\documentclass{IEEEcsmag}

\usepackage[colorlinks,urlcolor=blue,linkcolor=blue,citecolor=blue]{hyperref}

\usepackage{upmath}
\usepackage{amsthm}
\usepackage{stfloats}
\usepackage{placeins}
\usepackage{afterpage}
 
\newtheorem{definition}{Definition}

\jvol{XX}
\jnum{XX}
\paper{8}
\jmonth{March}
\jname{Accepted in the Applications department for the May/June 2025 issue of IEEE CG\&A}
\pubyear{2025}

\begin{document}

\sptitle{DEPARTMENT: UBIQUITOUS VIRTUAL REALITY}

\title{Meta-Objects: Interactive and Multisensory Virtual Objects Learned from the Real World for Use in Augmented Reality}

\author{Dooyoung Kim}
\affil{KAIST, Daejeon, 34141, South Korea}

\author{Taewook Ha, Jinseok Hong, Seonji Kim, Selin Choi, Heejeong Ko}
\affil{KAIST, Daejeon, 34141, South Korea}

\author{Woontack Woo}
\affil{KAIST, Daejeon, 34141, South Korea}

\markboth{UBIQUITOUS VIRTUAL REALITY}{UBIQUITOUS VIRTUAL REALITY}

\begin{abstract}
We introduce the concept of a meta-object, a next-generation virtual object that inherits the form, properties, and functions of its real-world counterpart, enabling seamless synchronization, interaction, and sharing between the physical and virtual worlds. While plenty of today’s virtual objects provide some sensory feedback and dynamic behavior, meta-objects fully integrate interactive and multisensory features within a structured data framework to enable real-time immersive experiences in a post-metaverse intelligent simulation platform. Three key components underpin the utilization of meta-objects in the post-metaverse: property-embedded modeling for physical and action realism, adaptive multisensory feedback tailored to user interactions, and a scene graph-based intelligence simulation platform for scalable and efficient ecosystem integration. By leveraging meta-objects through wearable AR/VR devices, the post-metaverse facilitates seamless interactions that transcend spatial and temporal barriers, paving the way for a transformative reality-virtuality convergence.
\end{abstract}


\maketitle

\chapteri{T}he widespread adoption of Augmented Reality/Virtual Reality (AR/VR) head-mounted displays (HMDs) has ushered in the era of ubiquitous virtual reality~\cite{lee2008recent}, blending the digital and physical realms to create new forms of interaction and communication\footnote{https://www.apple.com/apple-vision-pro/}\footnote{https://www.meta.com/kr/en/quest/quest-3/}.
At the heart of this evolution lies the concept of the metaverse, where virtual and real worlds interconnect to offer deeper and richer user experiences~\cite{shin2024evaluating}.
Within the metaverse, individuals can engage in interactions that transcend physical limitations, making the virtual living experience almost indistinguishable from reality, achieving what we call a \textbf{post-metaverse}. The \textbf{post-metaverse} is a reality-virtuality convergence economic platform implemented using wearable AR/VR devices. One of the components essential to achieving an immersive experience in the post-metaverse is to make users feel as if they are interacting with real objects, even though they are engaging with virtual ones.
Furthermore, for a virtual object to be truly ready for the post-metaverse, the changes resulting from a user's interactions in the virtual environment must also be reflected in its real-world counterpart. One example scenario involves a user controlling a virtual drone to transport two different boxes—a metallic one and a plastic one—each with distinct physical properties. As the drone lifts and moves the boxes, adaptive haptic feedback simulates the weight, texture, and temperature differences, while a real drone mirrors these actions miles away in real-time.

Despite advancements in AR/VR technologies, existing virtual objects remain fundamentally limited in their ability to provide immersive and realistic user experiences.
Currently, existing virtual objects are predominantly limited to reproductions of geometric forms and surface textures as rendered on desktop computers.
As a result, they exist merely as static entities, lacking the essential physical and action properties that would enable users to perceive them as real-world counterparts. 
The absence of these attributes restricts the expression of these objects, which further exacerbates interactions through wearable devices such as HMDs, leading to “shallow” and “devoid” experiences.
Moreover, existing interaction methods of wearable AR/VR devices fail to support fine-grained manipulation and complex operations on virtual objects~\cite{tong2023survey}.
Additionally, current feedback systems rely on a narrow range of sensory modalities, highlighting the need for sophisticated multi-modal approaches to achieve realistic and multisensory experiences~\cite{deng2023phyvr}.
Furthermore, interactions between physical and virtual worlds are often unidirectional—where only the physical influences the virtual—while reverse interactions remain limited.
Such constraints prevent the realization of seamless synchronization between the two realms, impeding the realization of interconnected, indistinguishable, and immersive experiences of the metaverse.
\begin{figure*}[ht]
 \centering
 \includegraphics[width=\linewidth]{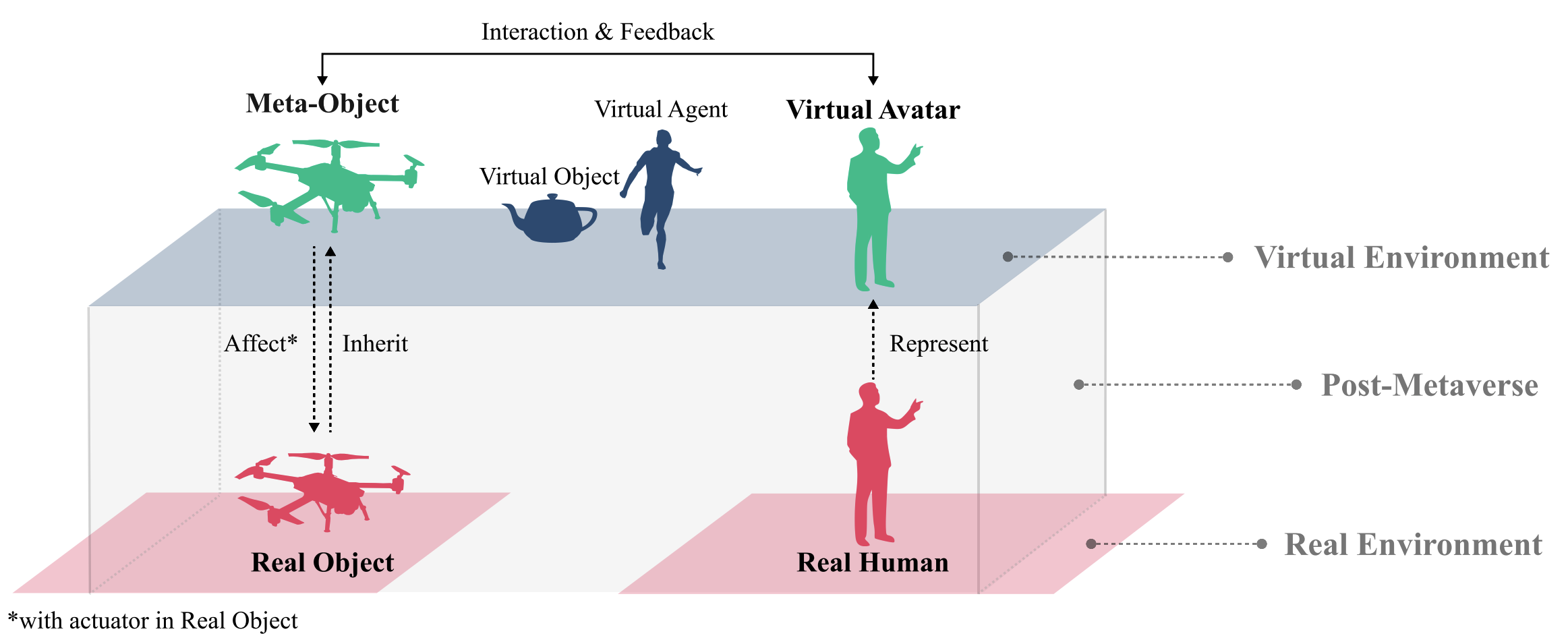}
 \caption{The difference between a meta-object and a virtual object is that it inherits the form, properties, and functions from a corresponding physical object, and it can affect the counterpart physical object. The virtual avatar of the real user could manipulate the real object in remote reality by interacting with a meta-object in the metaverse.}
 \label{fig:concept}
\end{figure*}

To address the limitations above, in this paper we present a conceptual framework for what we call a \textbf{meta-object}, a virtual entity that inherits the form, properties, and functions of its real-world counterpart, enabling seamless synchronization, interaction, and sharing between the physical and virtual worlds.
As illustrated in \autoref{fig:concept}, meta-objects differ from existing virtual objects in that they not only inherit properties and functions from physical objects but also influence corresponding physical entities with actuators in the physical environment.
To this end, a meta-object is characterized by key components: 1) physical and action properties inherited from the real-world counterpart, 2) interaction-adaptive multisensory feedback through wearable AR/VR HMDs and haptic devices, and 3) a scene graph-based data structure for adaptive property rendering, which dynamically adjusts feedback and rendered properties based on the capabilities of the output device.
In this paper, we will outline a comprehensive framework to demonstrate how the post-metaverse should be developed to integrate meta-objects effectively from three perspectives: 1) creation, 2) experience, and 3) an economic system.

\section{BACKGROUND}

\subsection{Extending Today's Virtual Objects}
To reproduce lifelike experiences in the metaverse, virtual reality systems attempt to replicate and mimic virtual objects both visually and physically to resemble their real-world counterparts.
To support visually realistic reconstruction, most systems have focused on rendering and modeling techniques to achieve high-quality visual representations of real objects.
In recent years, research in virtual object modeling has advanced through various deep learning techniques for 3D model generation, manipulation, and reconstruction~\cite{kerbl20233d}. 
Nontheless, off-the-self virtual objects are generally limited to being modeled as rigid body parts in 3D, allowing only for basic geometric manipulations such as movement, assembly, and size adjustments\footnote{https://learn.microsoft.com/en-us/windows/mixed-reality/mrtk-unity/mrtk3-overview/}.
Even with state-of-the-art visual reconstruction techniques, creating realistic 3D objects requires experts to spend a lot of time using specialized tools, causing limited reproducibility~\cite{durranicomplete}.

To represent and mimic the physical properties of objects, studies on haptic feedback encompass considerations of weight, temperature, and pressure, as well as tactile properties, such as slippery or rough textures. 
Still, modeling objects in a way that inherently embeds such properties and enables more diverse interactions with users remains underexplored.
This underscores the necessity of extracting, learning, and embedding real-world properties while creating virtual objects. 
Beyond visuals, these 3D models integrate properties such as the object's category, model, and required actions within 3D object datasets~\cite{yang2022oakink}.
Yet, current 3D models have primarily been used for simple labeling and elucidating object internals.
To convey realistic physical responses during user interactions, it is essential to incorporate physical properties and action properties that reflect real-world dynamics.

As virtual objects contain more detailed information to represent physical properties, actions, and user behaviors, they inevitably form complex and comprehensive data structures.
Moreover, relational data structures become necessary to handle user interactions and topological changes within objects ($e.g.$, disassembly, reassembly, and deformation). 
In fields including spatial perception, 3D reconstruction, and robotics, such comprehensive and relational data have been managed through a hierarchical structure, which allows efficient management of large volumes of data within the platform.

\subsection{Extending Today's Interaction and Feedback}\label{subsec2-2}
Interaction with virtual objects is currently limited due to simple hand gesture-based interaction systems and primitive sensory feedback that don't account for the properties of virtual objects.
Commercial HMDs rely on a few basic hand gestures, such as pinching, grabbing, and poking, due to unstable hand-tracking performance.
While hand-tracking performance has improved in recent years, it still has limitations when dealing with rapid movements, restricted tracking areas, and high computational requirements.
These limitations have led commercial HMDs to adopt simple gesture interactions rather than more dexterous hand interactions. 
To address these limitations, multi-modal tracking systems have been widely adopted for more stable and satisfactory performance. In multi-modal tracking systems, sensors with different modalities, such as inertial measurement units~\cite{streli2023hoov} and surface electromyography~\cite{gao2021hand}, are integrated with vision sensors, such as RGB+depth cameras. Through such sensor fusion methods, it is possible to implement not only stable interaction systems but also cost-efficient systems that can be integrated into AR/VR HMDs.

To provide realistic experiences in a reality-virtuality converged metaverse, it is crucial to process user interactions as input and generate adaptive output feedback accordingly.
While high-end hardware for display, spatial sound, and haptic feedback would be ideal, it is more practical to implement an efficient sensory feedback system that can be accomplished with fewer sensors and actuators and lower computational power within current hardware capabilities.
For this purpose, multisensory feedback that enhances sensory experiences by combining these three types of stimuli appears to be a promising approach.
Utilizing multiple stimuli for feedback has been confirmed to increase immersion and presence in virtual environments.
There have been studies on multisensory feedback that combine two types of feedback - visual and haptic - into a single mechanism~\cite{williams2020exploring}, and further results have demonstrated that incorporating all three stimuli - visual, auditory, and tactile - can more effectively enhance immersion and user experience compared to two-sense feedback~\cite{marucci2021impact}. A recent study~\cite{melo2020multisensory} has highlighted the importance of an advanced multisensory approach.
Therefore, integrating multisensory feedback is essential for reproducing realistic experiences and enhancing user engagement in the metaverse.

\subsection{Extending Today's Metaverse Platforms}\label{subsec2-3}
To move beyond conventional metaverse experiences, it is essential not only to extend virtual objects into meta-objects by embedding new properties and structures but also to evolve current metaverse platforms into the post-metaverse, where the physical and virtual worlds are seamlessly connected.
Generative AI techniques and models have the potential to enable users to automatically and effortlessly generate and create virtual objects within metaverse platforms.
However, these techniques often fail to accurately capture or learn context-aware configurations, particularly when factors such as user behaviors are involved.
To address this, the creation process should incorporate human-in-the-loop structures and approaches, allowing multiple users to actively participate in the generation phase.
Despite this need, many metaverse platforms face challenges in the creation process due to low accessibility for users.
To fulfill the role of the metaverse platforms, the platform should facilitate creation in a way that encourages participation from a wide range of users.

The economic system of the metaverse plays a crucial role in resolving user engagement, with multiple studies confirming its strong connection to sustaining user participation.
Key factors influencing user participation include reward mechanisms in the early stages and social behavioral environments in later stages in the metaverse ecosystem.
Reward systems in the metaverse often leverage Non-Fungible Tokens (NFTs) and blockchain-based digital assets, incorporating and utilizing digital asset trading systems to enhance user participation.
Through these systems, users can create digital content and activate the trading ecosystem within the metaverse platform~\cite{huawei2023economic}.
Therefore, an economic system can effectively incentivize participation, encouraging user engagement in the metaverse.


\section{KEY COMPONENTS OF META-OBJECTS}


\begin{figure*}[ht]
 \centering
 \includegraphics[width=\linewidth]{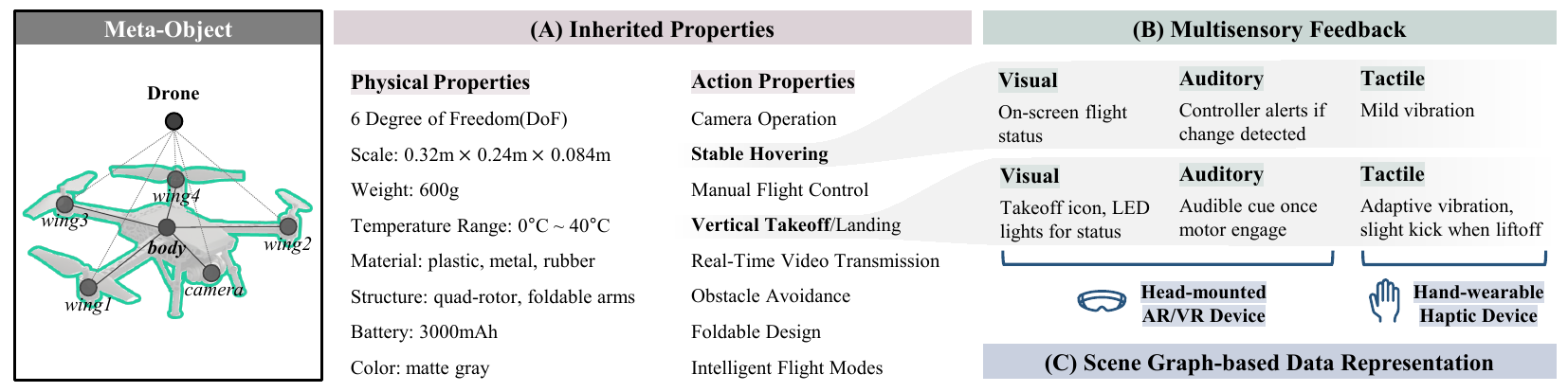}
 \caption{Three key components of a meta-object with a virtual drone example: (A) Inherited properties, (B) multisensory feedback, and (C) scene graph-based data representation.}
 \label{fig:metaobject}
\end{figure*}

In a metaverse that integrates the real and virtual worlds, meta-objects must acquire and incorporate information about their corresponding real-world counterparts, as well as user interaction data from reality, in order to achieve a seamless fusion of the physical and digital domains. We define a meta-object as follows:

\begin{definition}[Meta-Object]
A \emph{\textbf{meta-object}} is a virtual entity that inherits the form, properties, and functions of its real-world counterpart, enabling seamless synchronization, interaction, and sharing between the physical and virtual worlds.
\end{definition}

From an object-oriented perspective, meta-objects operate much like digital twins~\cite{jones2020characterising}, enabling seamless synchronization between the physical and virtual worlds and allowing real-world objects to interact dynamically with their digital counterparts in the metaverse. They are created based on real-world objects, ensuring a seamless linkage and synchronization between the two realms. These virtual entities inherit the shape, function, and intrinsic properties of their physical origins, allowing users to experience a sense of familiarity and coherence. Moreover, meta-objects are designed to be interactive, responding dynamically to user input and environmental changes in the virtual space. Importantly, these interactions are not one-sided; changes made in the virtual representation can be reflected back into the physical world, enabling real-time feedback and adaptive functionality. This bidirectional interaction enhances immersion, usability, and the integration of digital elements into everyday life, making meta-objects a key component in the evolving metaverse. The following sections will elaborate on the three key elements of meta-objects: inherited properties, multisensory feedback, and scene graph-based data representation.


\textbf{Inherited Properties}: Meta-objects integrate both \textit{physical} and \textit{action} properties to deliver authentic, interactive experiences in virtual and augmented environments. Their physical properties—including scale, weight, temperature range, material, structure, etc.—are learned from a physical object and embedded in a meta-object (see \autoref{fig:metaobject}(A) Physical Properties). This level of material fidelity ensures realistic behavior, such as proper weight distribution or deformation under pressure, providing users with immersive tactile feedback. By faithfully embedding these characteristics, meta-objects seamlessly bridge the gap between virtual and real worlds.

In addition to material fidelity, meta-objects incorporate action properties like disassembly, assembly, transformation, and destruction, enabling diverse and context-aware interactions (see \autoref{fig:metaobject}(A) Action Properties). Designed with interconnected components and movable parts, meta-objects adapt their responses based on how users handle or manipulate them. This allows for dynamic reactions—such as movement, structure changing, or simulated destruction—mirroring complex real-world behaviors. As a result, users experience meta-objects not just as visually accurate models but as interactive entities capable of evolving and responding naturally within reality-virtuality converged spaces.

\textbf{Multisensory Feedback}: Meta-objects in the metaverse capture nuanced hand movements to deliver context-sensitive, realistic interactions. Rather than recognizing only a single “grab” action, they track each finger’s contact point for varied responses, even with similar gestures. By integrating vision-based methods and a minimal set of sensors, systems interacting with meta-objects can detect subtle differences in hand posture or force, allowing the objects to adapt their shape or behavior accordingly. This ensures highly responsive and efficient experiences suitable for everyday use on wearable AR/VR HMDs—such as HoloLens 2, Quest 3, Vision Pro, or Varjo XR 4—alongside a haptic device for tactile engagement.

Building on this responsiveness, meta-objects employ integrated visual, auditory, and tactile feedback to heighten immersion without overburdening computational or energy resources. Rather than relying on numerous or costly sensors, the system uses a minimal actuator-based haptic device that delivers texture, force, and basic thermal cues. \autoref{fig:metaobject}(B) shows examples of the multisensory feedback from the meta-object (drone) that can result from taking various actions. During vertical takeoff, the system provides real-time sensory feedback through visual cues such as LEDs and a takeoff icon, auditory cues from the motor's sound, and haptic feedback simulating the liftoff sensation, enabling a highly realistic control and feedback experience. Combined with synchronized visuals and spatial audio tuned to each meta-object’s material properties, these complementary sensory channels augment the user’s perception and sense of presence, resulting in seamlessly lifelike interactions that transcend the limitations of traditional AR/VR systems.

\textbf{Scene Graph-based Data Representation:} Meta-objects employ a scene graph-based data model composed of nodes and edges to manage their physical properties, functional capabilities, and interaction data. Each node represents subparts, characteristics, or interaction histories, while edges define structural, hierarchical, and semantic connections. As shown in \autoref{fig:metaobject}'s drone example, a meta-object can contain a root node with child nodes for its base, wings, camera, and other sensors, all reflecting relationships between them. This design supports precise tracking of user interactions and a clear understanding of how individual subparts relate to one another, facilitating intelligent simulations in real-time virtual environments.

Beyond subpart relationships, meta-objects need to adapt to each user’s hardware setup, network conditions, and personal context—particularly in metaverse scenarios where wearables, XR devices, and connection speeds can vary widely. A meta-object’s graph-based nature allows it to selectively enable or disable certain nodes and properties depending on device capabilities, ensuring consistent yet optimized experiences across 3D stereo HMDs and audio systems ranging from basic to full spatial audio. As shown in \autoref{fig:metaobject}(C), tactile feedback nodes might be active only for users with haptic controllers, while visual nodes adjust resolution and audio nodes decide the level of details based on the headset’s specifications. This adaptability enables meta-objects to integrate physical and functional data for seamless collaboration across diverse devices and networks.

\section{BUILDING A POST-METAVERSE}

\begin{figure}[ht]
 \centering
 \includegraphics[width=\linewidth]{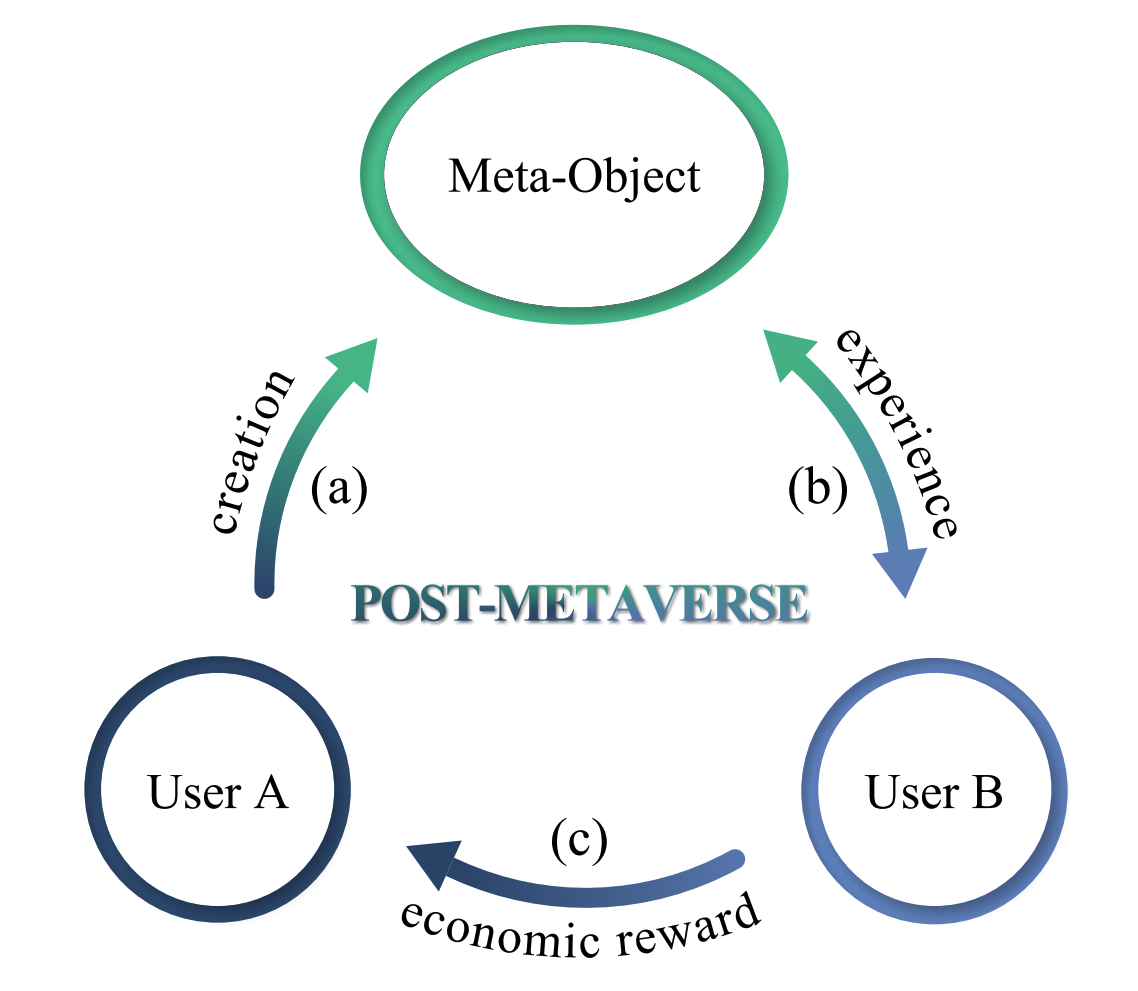}
 \caption{The interaction cycle between users and meta-objects in the post-metaverse. (a) User A (creator) creates a meta-object, (b) User B (consumer) experiences this meta-object, and (c) the economic rewards flow back to User A.}
 \label{fig:diagram}
\end{figure}


We have proposed the meta-object as a next-generation virtual object for enabling users to experience a reality-virtuality converged metaverse within physical spaces through wearable AR/VR devices. The meta-object is positioned as a foundational unit that enables participants to communicate, collaborate, and co-create in contexts that blur distinctions between virtual and real environments. Analyzing and expanding from the definition of metaverse-as-platform~\cite{shin2024evaluating} and the Ubiquitous Virtual Reality~\cite{lee2008recent}, we defined the platform for experiencing meta-object as follows:

\begin{definition}[Post-Metaverse]
The \emph{\textbf{post-metaverse}} is a reality-virtuality convergence economic platform that transcends the constraints of space and time to connect, communicate, and collaborate with people.
\end{definition}

\noindent The post-metaverse is conceptualized as a “reality-virtual convergence economic platform” that transcends the physical constraints of space and time, connecting individuals while enabling the sustainable production and consumption of meta-objects—which integrate interactive and multisensory characteristics learned from the real world. To achieve this, it is desirable (1) to apply digital twin technologies that accurately reflect real-world data, thereby facilitating the linkage and expansion of meta-objects; (2) to offer a user-friendly platform with technologies enabling more convenient and efficient meta-object creation and authoring; (3) to establish an NFT-based economic ecosystem that ensures the trustworthiness and interoperability of meta-objects, creating a secure environment for diverse stakeholders to maintain, manage, and expand digital assets; and finally, (4) to leverage wearable AR/VR devices so that users can fully experience the meta-objects’ multisensory qualities, delivering a rich user experience that seamlessly bridges reality and the virtual realm.

The meta-objects and users, who are the subjects of the post-metaverse, have an organic relationship, which can be expressed as a cycle in \autoref{fig:diagram}. In \autoref{fig:diagram}(a), the creation process in which an arbitrary user, represented as \textit{User A}, creates a meta-object including physical and action properties. The created meta-object exists in the post-metaverse and is consumed by a random user expressed as \textit{User B} in \autoref{fig:diagram}. At this time, \textit{User B} receives multisensory feedback from the meta-object and experiences an immersive and realistic \autoref{fig:diagram}(b) experience. This experience motivates \textit{User B} to consume the meta-object. At the same time, \textit{User B} pays a sufficient economic cost, which is ultimately returned to the creator, \textit{User A} as an economic reward in the post-metaverse. This reward motivates \textit{User A} to create the meta-object.

\subsection{Creation}

The creation of meta-objects in the post-metaverse begins with the generation of interactive, mesh-based models utilizing cutting-edge computer vision techniques. Methods such as Neural Radiance Fields (NeRF) for photorealistic reconstruction, 3D Gaussian Splatting for efficient point cloud rendering, and Mesh R-CNN for automated mesh extraction provide robust pipelines for producing high-quality digital representations of real-world objects. These technologies enable the accurate capture of object geometry, textures, and fine details while ensuring compatibility with interactive applications. By leveraging a learning-based approach, an object's parametric characteristics and interaction properties—derived from user behavior—are accurately modeled. Additionally, the Large Behavior Model (LBM) captures user behavior patterns, enabling the automatic generation of interactive properties in meta-objects. These approaches create scalable, realistic meta-object models that dynamically respond to user interactions in AR/VR. They streamline modeling for experts while making interactive 3D object creation more accessible for novices.



Once the foundational model is created, authoring tools allow users to enhance and refine the meta-object by embedding physical and action properties. Through intuitive interfaces, users with AR/VR devices can assign physical attributes that are lacking from initial creation to bring the digital entity closer to its real-world counterpart. These tools also enable embedding functions, such as how the object reacts to disassembly, transformation, or force application, ensuring contextual adaptability. By allowing general users to easily access, annotate, and refine meta-objects, the system ensures authenticity and relevance, making the virtual entities more meaningful and versatile.


To assign uniqueness and value to meta-objects, each meta-object is tokenized as a unique NFT, registering its provenance, properties, and creator details immutably. This ensures that creators retain intellectual property rights and can receive ongoing rewards or royalties whenever their meta-objects are utilized or traded. The NFT-based system not only secures ownership but also fosters a transparent and collaborative ecosystem where creators are incentivized to continuously innovate and refine their work. By integrating blockchain mechanisms, the post-metaverse ensures that meta-objects remain valuable, trustworthy, and interoperable across multiple platforms and applications, including AR/VR devices, completing the cycle of creation and sustaining the ecosystem.

\begin{figure*}[h]
 \centering
 \includegraphics[width=0.9\linewidth]{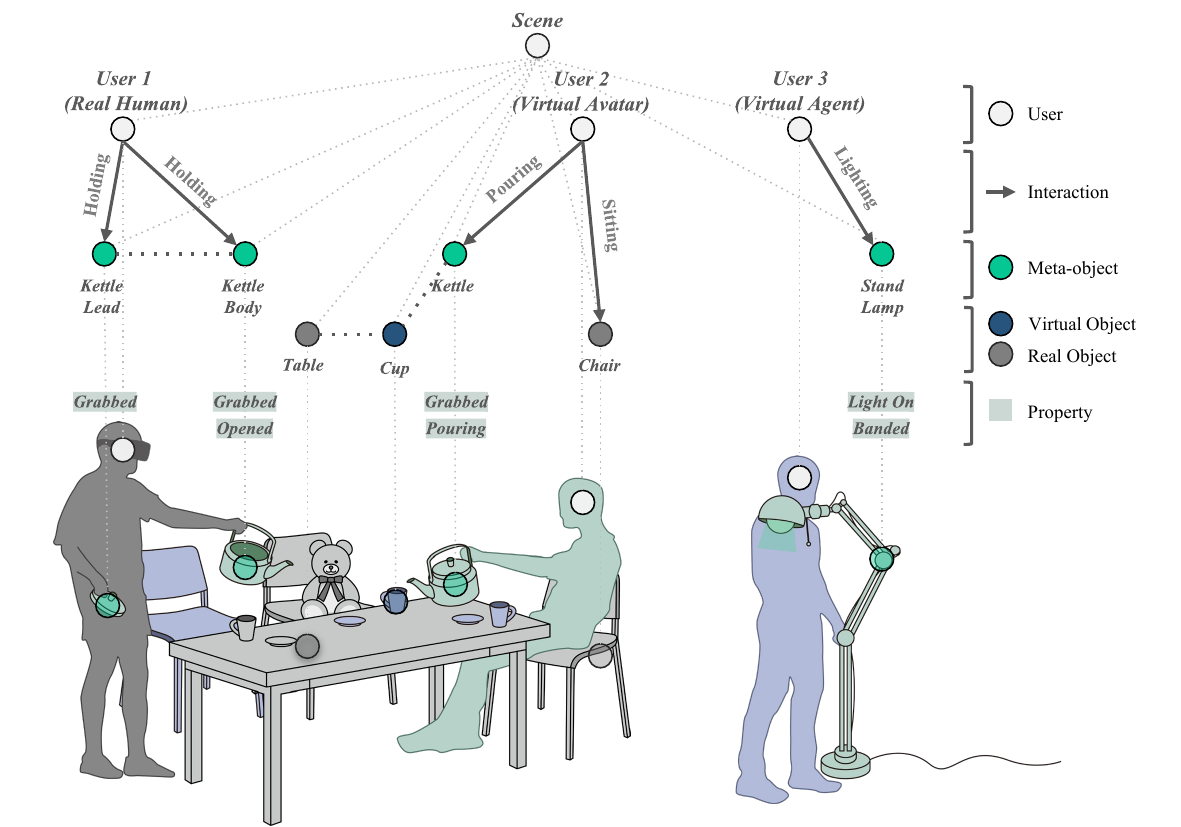}
 \caption{An overview of the use of meta-objects in the post-metaverse. Gray nodes represent real objects, green nodes represent meta-objects, and blue nodes represent virtual objects. Edges between objects represent semantic relationships, and directed edges between users and objects represent interactions.}
 \label{fig:experience}
\end{figure*}

\subsection{Experience} 

\begin{figure*}[h]
    \centering
    \includegraphics[width=\textwidth]{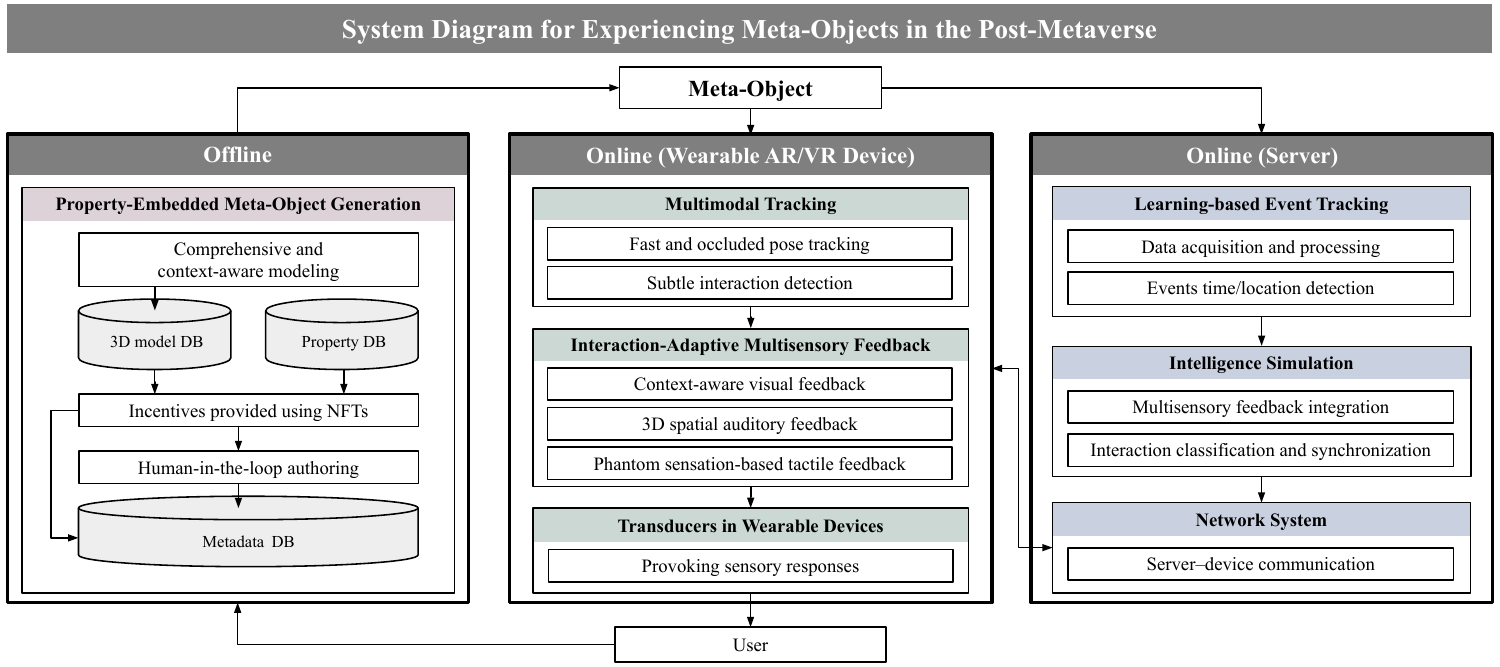}
    \caption{The system diagram for experiencing meta-object in post-metaverse.}
    \label{fig:post-metaverse_diagram}
\end{figure*}

\autoref{fig:experience} shows an overview of the use of meta-objects in the post-metaverse. In the post-metaverse, real humans wearing AR glasses, users accessing as virtual avatars via MR/VR headsets, and AI-powered virtual agents can collaboratively engage with meta-objects for immersive experiences. In the post-metaverse, users gain a sense of reality and engage in activities through their experiences with meta-objects. To provide immersiveness, interaction and feedback should be accomplished in real-time through wearable AR/VR devices, smart watches, haptic gloves, etc. This requires a system that is not only robust but also efficient. Their interactions are managed in real-time through a scene graph, ensuring synchronized responses and seamless interaction across spatial and temporal boundaries. Wearable devices and AR/VR HMDs provide visual and sensor data, enabling precise tracking of hand gestures and movements. As illustrated in \autoref{fig:post-metaverse_diagram}, the integration of a scene graph-based data structure and efficient computing resource allocation between HMDs and servers facilitates real-time capture of fine motor actions and rapid movements, enhancing the realism and responsiveness of the shared virtual environment.

To achieve an experience with meta-objects that equals or surpasses reality, multisensory feedback must be seamlessly integrated. To provide appropriate and sufficient visual feedback through AR/VR, HMDs should be capable of delivering not only high-fidelity graphic rendering but also realistic representations of contact points, appropriate responses to interactions, and visual expressions of physical phenomena such as destruction and deformation. Realistic audio feedback in meta-objects requires acoustic modeling based on real-world acoustic characteristics. By analyzing how spatial properties influence soundscapes during interactions, these insights can be integrated into meta-object interactions to enhance auditory realism. Tactile feedback can be delivered through haptic gloves with actuators and motors providing texture, force, and thermal sensations. Phantom sensations provide intuitive tactile feedback without the inconvenience of wearing haptic gloves~\cite{chen2024understanding}. This approach could potentially eliminate the need for complex actuators and motors while still delivering realistic texture, force, and thermal sensations. When combined with visual and auditory cues, it creates an accessible and comprehensive multi-modal experience that can be easily integrated into daily use.

A scene graph-based integration system is essential to seamlessly synchronize interactions and multi-modal feedback between meta-objects, the environment, and users. The meta-object handles diverse data types, including data for multi-modal tracking and inherent properties. Tracking results are acquired from sensors across various devices, which must be mutually calibrated. Moreover, the system must be designed to select appropriate feedback based on interaction and context from inherent properties of meta-objects and the environment. Using scene graphs enables nodes to simultaneously possess multiple features, and facilitates information exchange between nodes, making it possible to manage and utilize necessary data according to the overall context. 


\subsection{Economic System} 

To ensure sustained user participation in the post-metaverse, it is crucial to establish an incentive structure that motivates engagement from both meta-object creators and consumers. Initially, the economy can be driven by entities that utilize early-stage meta-objects to operate businesses, providing financial and reputational incentives to creators while offering value-added services to consumers. This foundational system encourages creators to contribute high-quality meta-objects and motivates consumers to adopt these objects for practical use. For example, businesses leveraging meta-objects in applications like virtual retail or education can share revenue with creators, fostering a collaborative and mutually beneficial ecosystem. Through mechanisms like blockchain-based ownership verification and NFT registration, creators retain intellectual property rights and receive royalties or rewards for their contributions. This cycle establishes trust and incentivizes ongoing participation, laying the groundwork for a robust economic platform.

Over time, the post-metaverse economy must evolve into a decentralized, user-driven platform where individuals can monetize their ventures using meta-objects. By enabling users to create and operate businesses—such as virtual design studios, gaming experiences, or educational services—meta-objects become tools for generating income. A portion of this revenue is reinvested into the ecosystem, benefiting both the platform and the original meta-object creators. This circular economy ensures that the value generated within the system flows back to sustain and expand it, creating a self-perpetuating model of growth. Interoperability across multiple virtual and real-world environments further enhances the economy’s scalability, allowing users to trade, lease, or repurpose meta-objects seamlessly across different contexts. This dynamic and inclusive economic framework transforms the post-metaverse into a sustainable platform for innovation, creativity, and financial opportunity, ensuring long-term engagement and societal impact.

\section{SYSTEM DIAGRAM OF POST-METAVERSE}

\autoref{fig:post-metaverse_diagram} shows the system diagram for experiencing meta-object in post-metaverse. The system comprises one offline module dedicated to the property-embedded meta-object generation and two online modules: one for tracking and multisensory feedback on a wearable AR/VR device and the other for a learning-based event tracking and intelligence simulation. The offline meta-object generation module in \autoref{fig:post-metaverse_diagram} automatically creates an initial version of the meta-object through comprehensive modeling and context-aware modeling to produce 3D models, coupled with property databases generated via user behavior learning. Subsequently, a user-participation incentive system offers rewards in the form of NFTs, and the meta-data created during this process is assigned to the preliminary meta-object, resulting in the final creation of the meta-object.

\autoref{fig:post-metaverse_diagram}'s online (wearable AR/VR device) module operating in wearable AR/VR devices processes acquired data and distinguishes data to be processed on-device and data to be offloaded. Tracked pose and subtle interaction from the multimodal tracking module could provide interaction-adaptive multisensory feedback by combining visual, auditory, and tactile feedback with transducers in wearable devices. The online (server) module in \autoref{fig:post-metaverse_diagram}, functioning on the server, utilizes offloaded data from wearable sensors along with information from environmental sensors to perform more complex posture and event estimations with machine-/deep learnings. It also uses a scene graph to conduct intelligence simulations to provide proper multisensory feedback by classifying and synchronizing interactions. Finally, through network systems, the server communicates with the AR/VR device in real-time, to provide life-like interaction and multisensory feedback with meta-objects on the post-metaverse.

The three distinct online and offline modules described in \autoref{fig:post-metaverse_diagram}—including an offline meta-object generation module in which users actively participate in creating meta-objects—are utilized to generate these objects and provide adaptive multimodal feedback during user interactions. The information of the created meta-objects is then transmitted to the online (wearable AR/VR device) and online (server) modules, respectively. Subsequently, the online (wearable AR/VR device) module receives the results of the intelligence simulation from the online (server) through the network system. This module then synchronizes the properties of the meta-object and the individual characteristics of the user, and classifies various interactions to generate customized multisensory feedback, which is delivered to the user through transducers. This process enables users to consistently produce meta-objects through voluntary participation. It provides a lifelike experience in the post-metaverse, even with users who are physically distant, as if they are together in the real world. This capability opens new opportunities in education and industry by vividly recreating past experiences or enabling experiences that are otherwise risky or costly.

\section{CONCLUSION}


This study explores the foundational role of meta-objects in shaping the post-metaverse, highlighting their potential to redefine digital creation, interaction, and economic systems. By integrating advanced modeling techniques, meta-objects achieve a high degree of realism and adaptability, enabling dynamic, multisensory interactions that seamlessly bridge the physical and virtual worlds. The proposed scene graph-based architecture ensures real-time synchronization of interactions across heterogeneous devices, facilitating collaborative experiences beyond spatial and temporal constraints. Furthermore, our work underscores the importance of computational efficiency in AR/VR environments, leveraging optimized HMD-server communication to enhance responsiveness and scalability.

Looking ahead, meta-objects offer transformative possibilities across various domains. Their integration with blockchain-based ownership and decentralized economic models fosters sustainable engagement by incentivizing creators and consumers, promoting a self-sustaining digital economy. Beyond entertainment and social interaction, this framework has broader implications for addressing global challenges such as remote collaboration, urban planning, and resource distribution. By enabling scalable, inclusive, and persistent shared experiences, the post-metaverse provides a platform for innovation, creativity, and societal advancement, paving the way for the next generation of human-computer interaction.


\section{ACKNOWLEDGMENT}

This work was supported by Institute of Information \& communications Technology Planning \& Evaluation(IITP) grant funded by the Korea government(MSIT) (No. RS-2024-00397663), the National Research Foundation of Korea(NRF) grant funded by the Korea government(MSIT) (No. RS-2021-NR059444), and under the metaverse support program to nurture the best talents (IITP-2024-RS-2022-00156435) grant funded by the Korea government(MSIT).

\bibliography{reference}
\bibliographystyle{IEEEtran}

\begin{IEEEbiography}{Dooyoung Kim}{\,} is a senior researcher with the KI-ITC Augmented Reality Research Center (ARRC) at KAIST, Daejeon, Korea. Contact him at dooyoung.kim@kaist.ac.kr.
\end{IEEEbiography}

\begin{IEEEbiography}{Taewook Ha}{\,} is a Ph.D. student in the Graduate School of Culture Technology (GSCT) at KAIST, Daejeon, Korea. Contact him at hatw95@kaist.ac.kr
\end{IEEEbiography}

\begin{IEEEbiography}{Jinseok Hong}{\,} is a Ph.D. student in the Graduate School of Metaverse (GSMV) at KAIST, Daejeon, Korea. Contact him at jindogliani@kaist.ac.kr.
\end{IEEEbiography}

\begin{IEEEbiography}{Seonji Kim}{\,} is a Ph.D. student in the Graduate School of Culture Technology (GSCT) at KAIST, Daejeon, Korea. Contact her at seonji.kim@kaist.ac.kr.
\end{IEEEbiography}

\begin{IEEEbiography}{Selin Choi}{\,} is a Ph.D. student in the Graduate School of Culture Technology (GSCT) at KAIST, Daejeon, Korea. Contact her at selin.choi@kaist.ac.kr.
\end{IEEEbiography}

\begin{IEEEbiography}{Heejeong Ko}{\,} is a M.S. student in the Graduate School of Culture Technology (GSCT) at KAIST, Daejeon, Korea. Contact her at hj.ko@kaist.ac.kr.
\end{IEEEbiography}

\begin{IEEEbiography}{Woontack Woo} {\,} is a head of the GSMV, a professor at the GSCT, and the director of the KI-ITC ARRC at KAIST, Daejeon, Korea. Contact him at wwoo@kaist.ac.kr.
\end{IEEEbiography}

\end{document}